\documentclass[RNAAS,twocolumn]{aastex62}

\usepackage{amssymb}
\usepackage{amsmath}
\usepackage{graphicx}
\usepackage{dcolumn}
\usepackage{hyperref}
\usepackage{color,units}
\usepackage{aas_macros}
\usepackage{lineno}
\usepackage{xspace}
\usepackage{subfigure}
\usepackage{comment}
\usepackage{newtxtext,newtxmath}

\usepackage[normalem]{ulem} 

\newcommand{\red}[1]{{#1}}

\newcommand{\bilby}{\textsc{Bilby}\xspace}

\newcommand{\seobnre}{\textsc{SEOBNRE}\xspace}
\newcommand{\imrphenomD}{\textsc{IMRPhenomD}\xspace}
\newcommand{\imrphenomPv}{\textsc{IMRPhenomPv2}\xspace}

\newcommand{\nrsurfour}{\textsc{NRSur7dq4}\xspace}

\newcommand{\eventname}{GW190521\xspace}
\newcommand{\tenhz}{\unit[10]{Hz}\xspace}

\newcommand{\SPA}{School of Physics and Astronomy, Monash University, Clayton VIC 3800, Australia}
\newcommand{\OzGravMonash}{OzGrav: The ARC Centre of Excellence for Gravitational Wave Discovery, Clayton VIC 3800, Australia}
\newcommand{\HongKong}{Department of Physics, The Chinese University of Hong Kong, Shatin, N.T., Hong Kong}

\begin{document}

\title{
\eventname: orbital eccentricity and signatures of dynamical formation in a \\binary black hole merger signal
}

\author{Isobel Romero-Shaw}
    \email{isobel.romero-shaw@monash.edu}
\affiliation{\SPA}
\affiliation{\OzGravMonash}

\author{Paul D. Lasky}
\affiliation{\SPA}
\affiliation{\OzGravMonash}

\author{Eric Thrane}
\affiliation{\SPA}
\affiliation{\OzGravMonash}

\author{Juan Calder{\'o}n Bustillo}
\affiliation{\SPA}
\affiliation{\OzGravMonash}
\affiliation{\HongKong}

\begin{abstract}
Pair instability supernovae are thought to restrict the formation of black holes in the mass range $\sim \unit[50 - 135]{M_\odot}$. 
However, black holes with masses within this ``high mass gap'' are expected to form as the remnants of binary black hole mergers.
These remnants can merge again dynamically in densely populated environments such as globular clusters.
The hypothesis that the binary black hole merger \eventname formed dynamically is supported by its high mass.
Orbital eccentricity can also be a signature of dynamical formation, since a binary that merges quickly after becoming bound may not circularize before merger.
In this work, we measure the orbital eccentricity of \eventname.
We find that the data prefer a signal with eccentricity $e \geq 0.1$ at \tenhz to a non-precessing, quasi-circular signal, with a log Bayes factor $\ln{\cal B}=5.0$.
When compared to precessing, quasi-circular analyses, the data prefer a non-precessing, $e \geq 0.1$ signal, with log Bayes factors $\ln{\cal B}\approx2$.
Using injection studies, we find that a non-spinning, moderately eccentric ($e = 0.13$) \eventname-like binary can be mistaken for a quasi-circular, precessing binary. 
Conversely, a quasi-circular binary with spin-induced precession may be mistaken for an eccentric binary.
We therefore cannot confidently determine whether \eventname was precessing or eccentric.
Nevertheless, since both of these properties support the dynamical formation hypothesis, our findings support the hypothesis that \eventname formed dynamically. 
\end{abstract}

\section{Introduction}
The first and second observing runs of the Advanced LIGO~\citep{Aasi13} and Virgo~\citep{AdvancedVirgo} gravitational-wave observatories yielded ten observations of stellar-mass black-hole binaries~\citep{abbott16_gw150914_detection, abbott16_01BBH}, reported in their first gravitational-wave transient catalogue~\citep[GWTC-1;][]{GWTC-1}.
The question of how these binaries came to merge within the age of the Universe remains unanswered. 
Proposed formation channels typically fall into two categories: \textit{isolated}, in which two stars evolve side-by-side until they form black holes and coalesce \citep[see, e.g.,][]{Livio88, Bethe98, deMink10, Ivanova13, Kruckow16, deMink16}, and \textit{dynamical}, in which two black holes become bound due to gravitationally-driven interactions inside dense star clusters \citep[e.g.,][]{Sigurdsson93, PortegiesZwart99, OLeary05, Samsing13, Morscher15, Gondan17, Samsing17, Rodriguez18b, Randall17, Randall18, SamsingDOrazio18, Samsing18, Rodriguez18a, Fragione18, Fragione19b, Bouffanais19} and/or active galactic nuclei disks~\citep{Yang2019,McKernan2020, Grobner20}. 
Young star clusters may create something of a hybrid channel, with dynamical interactions \red{perturbing the evolution of primordial} stellar binaries, which evolve to make merging double compact objects~\citep{Ziosi2014,DiCarlo2019,Rastello2020}.

The component masses and spins of a black-hole binary can illuminate its formation history, as can its orbital eccentricity (e.g., \citep{Vitale15, Stevenson15bqa, Rodriguez16, Farr17, Fishbach17, TalbotThrane17}).
Information about these parameters can be extracted from the gravitational-wave signal. 
Both isolated evolution and dynamical formation can produce black-hole binaries with properties like those presented in {GWTC-1}, with component masses $m_1, m_2 \lesssim 50 \text{M}_\odot$, dimensionless component spins $a_1, a_2$ consistent with $0$, and eccentricities $e$ consistent with $0$ at \tenhz \citep{GWTC-1, RoSho19}. 
Dynamical formation is the preferred pathway for binaries with more extreme masses~\citep{Gerosa17, Rodriguez19, Bouffanais19, fragione20, Fragione2020}, isotropically distributed spin tilt angles~\citep{Rodriguez16, TalbotThrane17}, and non-zero orbital eccentricities~\citep{Zevin17, Rodriguez18b, Gondan18, Samsing18, Zevin18}. 

The mass distribution of black holes that form as the remnants of massive stars is thought to deplete between $\sim 50$ and $\sim \unit[135]{M_\odot}$ due to pair-instability supernovae \citep[PISN;][]{HegerWoosley02, Ozel10, Belczynski16, Marchant16, Fishbach17, Woosley17} \red{unless exotic physics is invoked~\citep{Sakstein2020}}.
The precise lower limit of the PISN mass gap is an area of active reseach; see \cite{Belczynski2020}, and references therein.
The remnants of binary black hole merger events can have masses within the PISN gap; see, e.g., \citep{abbott16_gw150914_detection,Fishbach17a,Chatziioannou2019,Kimball2019,Kimball2020}.
Second-generation mergers---where at least one of the binary components is a remnant of a previous merger, potentially within the mass gap---can occur in the high-density environments conducive to dynamical mergers  \citep{Gerosa17, Rodriguez19, Bouffanais19, fragione20}.
Prior to the detection of \eventname, no convincing evidence has emerged for hierarchical mergers~\citep{Fishbach17a,Chatziioannou2019,EccentricCWB19,Kimball2019,Kimball2020}.

Isolated binaries are thought to circularize efficiently, leading to negligible eccentricity close to merger \citep{Peters64, Hinder07}.
While it is possible that the late-inspiral eccentricity of field mergers can be increased by Kozai-Lidov resonance \citep{Kozai62, Lidov62} during three-body \citep{Silsbee16, Antonini17, Fishbach17a, Rodriguez18jqu, triplespin, Liu19} and four-body \citep{quadruples, quadruples2} interactions in the field, the relative rate of such events is expected to be small, assuming moderate progenitor metallicities and black-hole natal kicks \citep{Silsbee16, Antonini17, Rodriguez18jqu, triplespin, Liu19}.
In contrast, some dynamically-formed binaries merge so rapidly after becoming bound that they retain non-negligible eccentricity in the LIGO--Virgo band \citep{Zevin17, Rodriguez18a, Samsing18, Gondan18, Zevin19ns}. 
Multiple authors \citep[e.g.,][]{Samsing17, SamsingRamirez17, SamsingDOrazio18, Rodriguez18b, Rodriguez18a} show that we can expect $\mathcal{O}(5\%)$ of all dynamical mergers in globular clusters to have eccentricities $e > 0.1$ at a gravitational-wave frequency of \tenhz. 

The LIGO-Virgo Collaboration recently announced the detection of \eventname, a gravitational-wave signal from the merger of a black hole binary with component masses $m_1 = \unit[85^{+21}_{-14}]{M_\odot}$, $m_2 = \unit[66^{+17}_{-18}]{M_\odot}$~\citep{GW190521, GW190521_implications}.
\red{The median and 90\% confidence intervals quoted for these masses place at least one component within the PISN mass gap.\footnote{\red{\cite{Fishbach2020} find that $m_1$ is above the mass gap if $m_2 < \unit[48]{M_\odot}$, below the mass gap}.}}
The data exhibit a modest preference (log Bayes factor $\text{ln}~\mathcal{B} \approx 2.4$) for spin-induced precession of the orbital plane, suggesting that the black-hole spin vectors may be significantly misaligned from the orbital angular momentum axis.
If confirmed, the signature of precession would lend support for the dynamical hypothesis. 

\begin{figure*}
    \centering
    \begin{subfigure}
        \centering
        \includegraphics[width=0.35\linewidth]{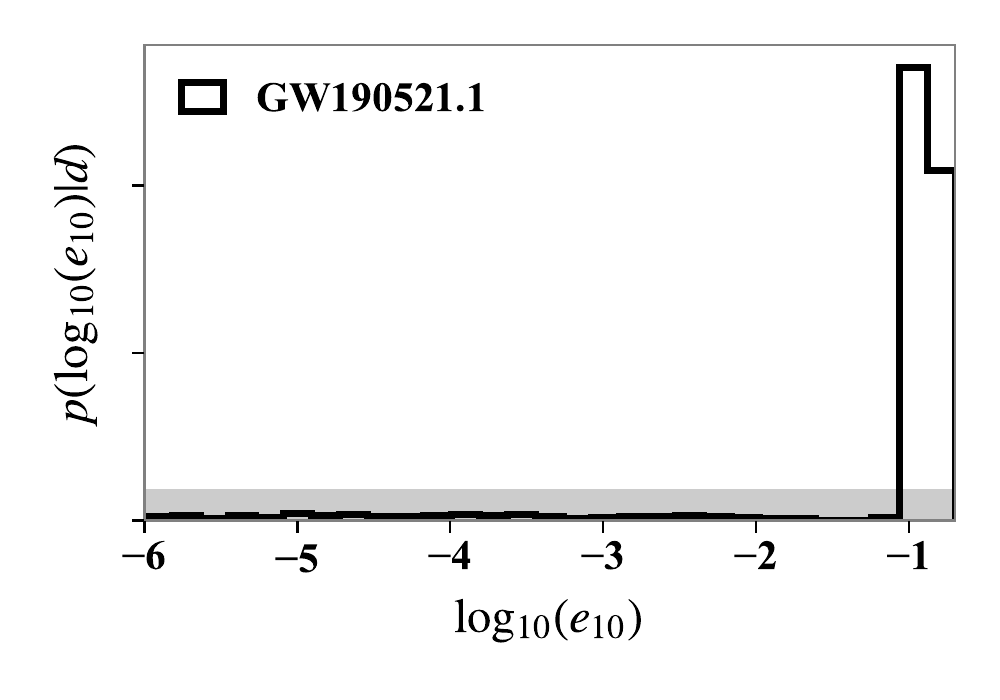}
    \end{subfigure}
    ~
    \begin{subfigure}
        \centering
        \includegraphics[width=0.35\linewidth]{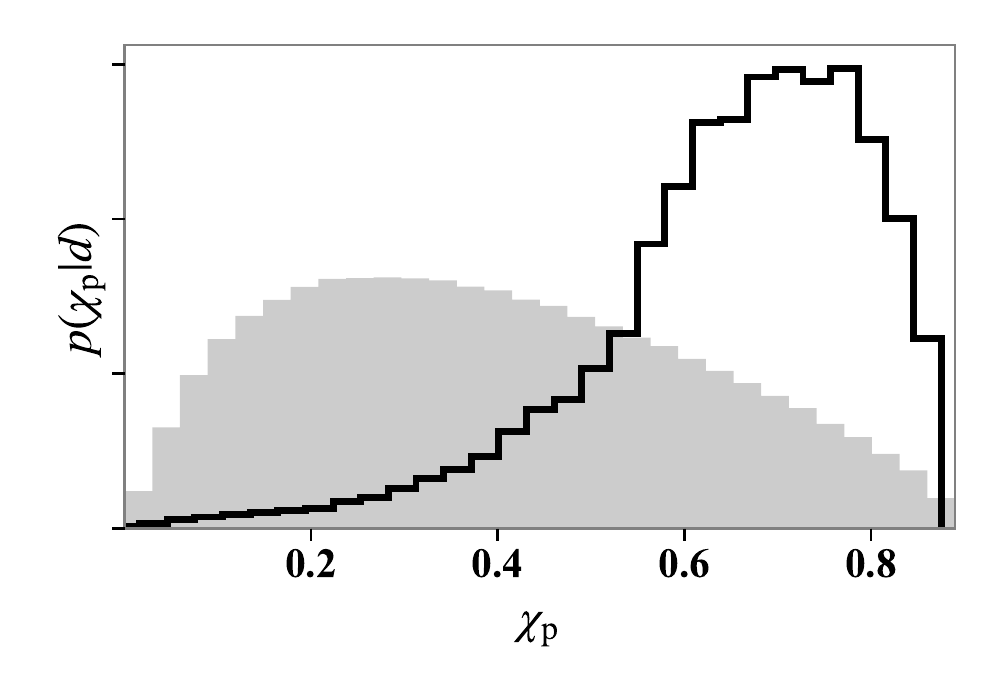}
    \end{subfigure}
    \caption{Results of analysis of \eventname using \seobnre and \imrphenomPv. \textit{Left}: posterior probability density distribution for eccentricity at \tenhz for \eventname, recovered using \seobnre. At 90\% confidence, $e_{10} \geq 0.11$. The posterior rails against the upper limit of the prior, $e_{10} = 0.2$, suggesting that the true value lies beyond this waveform-enforced constraint. \textit{Right}: posterior probability density distribution for the precession parameter $\chi_p$ for \eventname, recovered using \imrphenomPv. The prior probability for each parameter is shown in gray. \label{fig:eccentric_precessing_event}}
\end{figure*}

\begin{table*}[]
\caption{\label{tab:recovered_values_GW190521} Recovered \eventname parameter values obtained using eccentric waveform model \seobnre, precessing waveform models \imrphenomPv and \nrsurfour, and \nrsurfour constrained to have aligned spins. For the \seobnre analysis, we give the 90\% confidence lower limit on eccentricity at \tenhz. For other parameters, the median of the posterior is given along with the 90\% credible interval. \red{In the final column, we state the values inferred from the LIGO--Virgo analysis, read from the public posterior samples obtained using \nrsurfour~\citep{GWOSC}.} In the final row, we provide the log Bayes factor of each analysis against the signal-to-noise log Bayes factor obtained for $e_{10} \geq 0.1$ using \seobnre ($\mathrm{ln}~{\mathcal{B}_{S/N}} = 85.7$).}
\begin{tabular}{l||c|c|c|c|c}

Parameter (source frame) & \seobnre & \imrphenomPv & \nrsurfour & \nrsurfour aligned & \red{\nrsurfour LVC} \\
\hline \hline
Primary mass, $m_1$ [$\mathrm{M}_\odot$]  & $92^{+26}_{-16}$ & $126^{+61}_{-41}$ & $86^{+18}_{-13}$ & $85^{+22}_{-14}$ & \red{$85^{+21}_{-14}$}\\
Secondary mass, $m_2$  [$\mathrm{M}_\odot$] & $69^{+18}_{-19}$ & $59^{+32}_{-24}$ & $69^{+18}_{-17}$ & $61^{+15}_{-17}$  & \red{$66^{+17}_{-18}$}\\
Luminosity distance, $d_\mathrm{L}$ [Gpc] & $4.1^{+1.8}_{-1.8}$ & $2.4^{+2.3}_{-1.0}$ & $4.7^{+2.2}_{-2.2}$ & $4.7^{+1.6}_{-1.5}$ & \red{$5.3^{+2.4}_{-2.6}$} \\
Right ascension, $\alpha$ [rad] & $3.6^{+2.7}_{-3.5}$ & $4.3^{+1.9}_{-4.3}$ & $3.4^{+2.9}_{-3.4}$ & $3.7^{+2.6}_{-3.7}$ & \red{$3.5^{+2.8}_{-3.4}$}\\
Declination, $\delta$ [rad] & $-0.7^{+1.4}_{-0.5}$ & $-0.7^{+1.5}_{-0.4}$ & $-0.8^{+1.5}_{-0.4}$ & $-0.9^{+1.6}_{-0.3}$  & \red{$-0.8^{+1.5}_{-0.4}$}\\
Reference phase, $\phi$ [rad] & $3.1^{+2.9}_{-2.7}$ & $3.0^{+3.0}_{-2.7}$ & $3.2^{+2.6}_{-2.6}$ & $3.1^{+2.9}_{-2.8}$ & \red{$3.4^{+ 2.6}_{-3.2}$}\\
Polarisation, $\psi$ [rad] & $1.5^{+1.5}_{-1.4}$ & $1.6^{+1.3}_{-1.5}$ & $1.8^{+1.2}_{-1.5}$ & $1.6^{+1.4}_{-1.4}$ & \red{$1.8^{+1.2}_{-1.6}$}\\
Inclination, $\theta_\mathrm{JN}$ [rad] & $1.3^{+1.6}_{-1.0}$ & $1.4^{+1.0}_{-0.7}$ & $0.8^{+2.0}_{-0.6}$ & $0.7^{+2.2}_{-0.5}$ & \red{$0.8^{+2.1}_{-0.6}$}\\
Eccentricity lower limit at \tenhz, $e_{10}^\mathrm{min}$ & $0.11$ & N/A & N/A & N/A & \red{N/A}\\
Effective spin, $\chi_\mathrm{eff}$ & $0.0^{+0.2}_{-0.2}$ & $0.1^{+0.4}_{-0.4}$ & $0.0^{+0.3}_{-0.3}$ & $0.0^{+0.2}_{-0.3}$ & \red{$0.1^{+0.3}_{-0.4}$}\\
Effective precession, $\chi_p$ & N/A & $0.7^{+0.2}_{-0.3}$ & $0.6^{+0.2}_{-0.3}$ & N/A & \red{ $0.7^{+0.3}_{-0.4}$}\\
Log Bayes factor against \seobnre, $\mathrm{ln}~{\mathcal{B}_{X/E}}$ & $0.0$ & $-2.0$ & $-1.8$ & $-5.0$ & \red{$-1.2$}
\end{tabular}
\end{table*}

In this work, we show that \eventname is consistent with an eccentric merger.
For brevity, we hereafter refer to the eccentricity measured at a gravitational-wave frequency of \tenhz as $e_{10}$.
Our method allows us to study eccentricities up to $e_{10} = 0.2$\red{, beyond which the waveform is not available}.
Our analysis reveals overwhelming support for a spin-aligned eccentric signal with $e_{10} \geq 0.1$ over a spin-aligned quasi-circular signal. 
We use simulated events to demonstrate that precession and eccentricity cannot be distinguished for a \eventname-like signal.
We end with a discussion of the implications of our results on the potential formation mechanism of \eventname.

\section{Method}
We construct eccentric posterior probability density distributions using the method developed in \cite{RoSho19}, which is built on those introduced by \cite{Payne} and \cite{Lower18}. 
We use the Bayesian inference library \bilby~\citep{bilby, bilbyGWTC1} to perform an analysis using our ``proposal'' model: the spin-aligned quasi-circular waveform model \imrphenomD~\citep{Khan15}.
We reweight our \imrphenomD posteriors to our ``target'' model: the spin-aligned eccentric waveform \seobnre~\citep{SEOBNRE, validationSEOBNRE}. 
Our prior on eccentricity is log-uniform in the range $-6 \leq \log_{10}(e_{10}) \leq -0.7$.
\red{
The upper limit arises from waveform limitations, although even a model allowing higher eccentricities would be restricted by the reweighting method.
In order to reweight posterior samples efficiently, the samples obtained using the proposal model must cover the same region of the multidimensional parameter space as would be obtained by direct sampling with our target model.
The overlap between eccentric and quasi-circular waveforms with otherwise-identical parameters falls drastically for higher eccentricities, so their posterior samples would not reside in the same region of the parameter space.}
We marginalise over the time and phase of coalescence as in \cite{Payne} to account for differing definitions of these parameters between our proposal and target models.

\begin{table*}[]
\caption{\label{tab:injected_recovered_values} The 90\% credible upper limit on eccentricity at \tenhz, $e_{10}^\mathrm{max}$, and recovered precession parameter $\chi_p$ for different injections with varying waveform model, $e_{10}$ and $\chi_p$ settings. For the recovered $\chi_p$ we quote the posterior median and 90\% credible interval.}
\begin{tabular}{l||c|c|c|c}

Injected waveform model & Injected $e_{10}$ & Injected $\chi_p$ & Recovered $e_{10}^\mathrm{max}$ with \seobnre & Recovered $\chi_p$ with \imrphenomPv \\
\hline \hline
\imrphenomD         & 0                 & 0                 & 0.025                        & $0.39^{+0.37}_{-0.29}$                \\
\nrsurfour               & 0                 & 0                 & 0.032                                & $0.33^{+0.40}_{-0.25}$                \\
\seobnre                & 0                 & 0                 & 0.055                                & $0.42^{+0.36}_{-0.30}$                \\
\imrphenomPv            & 0                 & 0.63              & 0.077                                & $0.43^{+0.35}_{-0.32}$                \\
\nrsurfour               & 0                 & 0.63              & 0.118                                & $0.53^{+0.29}_{-0.37}$  \\
\seobnre                 & 0.13              & 0                 & 0.136                                & $0.57^{+0.26}_{-0.39}$                

\end{tabular}
\end{table*}

\begin{table}[]
\caption{\label{tab:injected_values} Parameters shared by all injected waveforms.}
\begin{tabular}{l|c}

Parameter (source frame) & Value  \\
\hline \hline
Primary mass, $m_1$ [$\mathrm{M}_\odot$] & $84$ \\
Secondary mass, $m_2$ [$\mathrm{M}_\odot$] & $62$  \\
Luminosity distance, $d_\mathrm{L}$ [Gpc] & $5.0$  \\
Right ascension, $\alpha$ [rad] & $3.3$ \\
Declination, $\delta$ [rad] & $0.5$ \\
Reference phase, $\phi$ [rad] & $6.2$ \\
Polarisation, $\psi$ [rad] & $1.6$ \\
Inclination, $\theta_\mathrm{JN}$ [rad] & $0.3$ \\
Geocent time, $t_0$ [s] & $1242442967.46$ \\
\end{tabular}
\end{table}

\section{Analysis of \eventname}
We analyze publicly-available data and noise power spectral densities from~\cite{GW190521, GWOSC}. We reproduce the settings of the LVC analysis for our parameter estimation, with a data segment of $\unit[8]{s}$, a frequency band $11$--$\unit[512]{Hz}$, and sampling frequency $\unit[1024]{Hz}$. In order to assess the role of waveform systematics, we perform four analyses using three different waveform models (one waveform is used twice with two different spin priors). The results of these analyses are summarized in Table \ref{tab:recovered_values_GW190521}.

First, we analyze the data using the aligned-spin eccentric waveform model \seobnre.
We present the posterior distribution on the $e_{10}$ of \eventname in the left-hand panel of Fig.~\ref{fig:eccentric_precessing_event}. 
The posterior drastically deviates from the log-uniform prior, strongly favouring eccentricities $e_{10} \geq 0.1$.
There is little support for $e_{10} < 0.1$, with $90\%$ of the posterior at $e_{10} \geq 0.11$. 
For other parameters, we obtain median posterior values similar to those given in Table 1 of \cite{GW190521}, with a median source-frame total mass $M=\unit[161^{+28}_{-20}]{M_\odot}$, mass ratio $q= 0.7^{+0.2}_{-0.3}$, and $\chi_\mathrm{eff}=0.0^{+0.2}_{-0.2}$. 
We obtain a luminosity distance, $d_\mathrm{L}=\unit[4.0^{+1.9}_{-1.7}]{Gpc}$, which is slightly lower than (but consistent with) the value of $\unit[5.3^{+2.4}_{-2.6}$]{Gpc} from the LIGO--Virgo analysis. 
Eccentricity causes a faster merger, reducing the signal power. Thus, in order to match the observed signal-to-noise ratio with an eccentric template, we may require a closer source. 
Additionally, models like \seobnre, \imrphenomD and \imrphenomPv, which do not contain higher-order modes, cannot rule out edge-on binaries, which reduces the median distance estimate~\citep{GW190521_implications}.
Posterior distribution plots for all other parameters are available online.\footnote{\href{https://git.ligo.org/isobel.romero-shaw/gw190521.1}{git.ligo.org/isobel.romero-shaw/gw190521.1}}

Next, we perform an analysis using the precessing waveform \imrphenomPv~\citep{IMRPhenomP} with otherwise-identical settings.
In Fig.~ \ref{fig:eccentric_precessing_event}, we show the posterior distribution for $\chi_p$ of \eventname obtained with \imrphenomPv.
This analysis recovers a smaller median $d_\mathrm{L}$ than the \seobnre analysis, with a more extreme mass ratio, $q \approx 0.5$.
In order to carry out model selection comparing the \imrphenomPv results to those obtained with \seobnre, we implement an astrophysically-motivated prior on eccentricity.
Theoretical studies robustly predict that $\sim5$\% of binaries that form dynamically in globular clusters will have $e_{10} \geq 0.1$~\citep[e.g.,][]{Samsing17, Samsing18, Rodriguez18a, Kremer2020}. To investigate this hypothesis, we assume a log-uniform distribution for $\log_{10} e_{10}\in (-1,- 0.7)$.
Using this astrophysically-motivated prior, the eccentric model is mildly preferred to the precessing model by a factor of $\mathrm{ln}~{\mathcal{B}_{E/P}} = 2.0$.
If we repeat the same calculation using the (less well-motivated) prior range $\log_{10}e_{10}\in(-6,-0.7)$ as in Fig.~\ref{fig:eccentric_precessing_event}, the eccentric (E) and precessing (P) waveform models are almost equally well-supported by the data, with a log Bayes factor $\mathrm{ln}~{\mathcal{B}_{E/P}} = -0.35$.

Finally, we perform computationally-intensive analyses using the precessing, higher-order-model waveform \nrsurfour, using parallel \bilby~\citep{ParallelBilby} to manage computational costs. We run two versions of the \nrsurfour analysis: one assuming aligned black-hole spins (no precession) and one allowing arbitrary spin orientations (allowing precession). Otherwise, the assumptions are identical to the \imrphenomPv analysis above. While the two \nrsurfour analyses obtain near-identical results, the analysis that includes precession (P) is preferred over the no-precession hypothesis with a moderate $\mathrm{ln}~{\mathcal{B}_{P/NP}} = 3.2$. The eccentric \seobnre hypothesis (with $e_{10}>0.1$) is preferred to the precessing and non-precessing \nrsurfour analyses by log Bayes factors of $\mathrm{ln}~{\mathcal{B}_{E/P}} = 1.8$ and $\mathrm{ln}~{\mathcal{B}_{E/NP}} = 5.0$, respectively.

We perform two additional analyses, identical in almost all aspects to the \nrsurfour studies described above, but without including higher-order modes. If we assume aligned spin, we obtain results similar to the \seobnre analysis. If we allow for precession, we obtain results similar to the \imrphenomPv results with luminosity distance $\unit[2.8^{+2.2}_{-1.5}]{Gpc}$ (90\% credibility)  and $q \approx 0.5$.


\begin{figure*}
    \centering
    \begin{subfigure}
        \centering
        \includegraphics[width=0.35\linewidth]{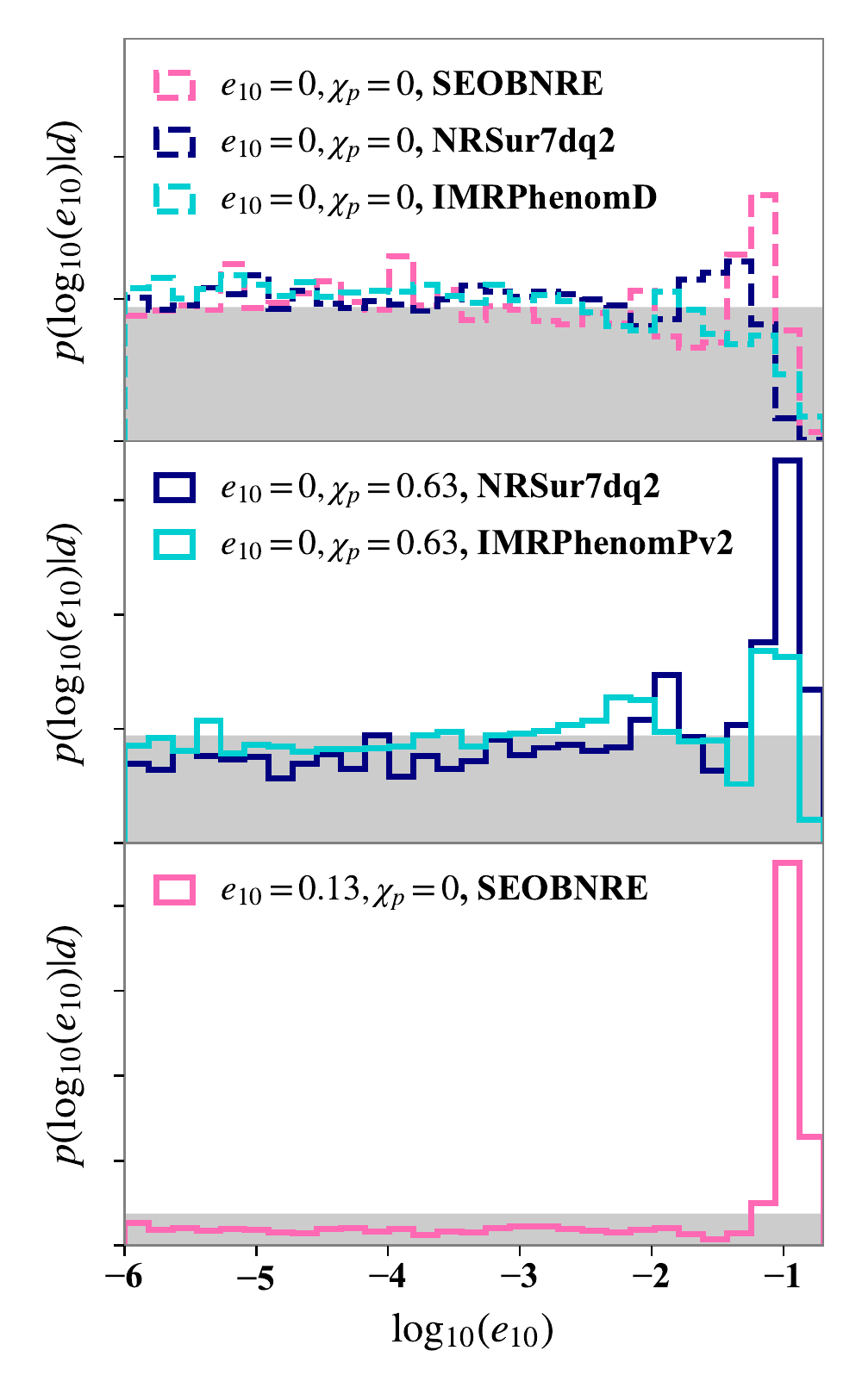}
    \end{subfigure}
    ~
    \begin{subfigure}
        \centering
        \includegraphics[width=0.35\linewidth]{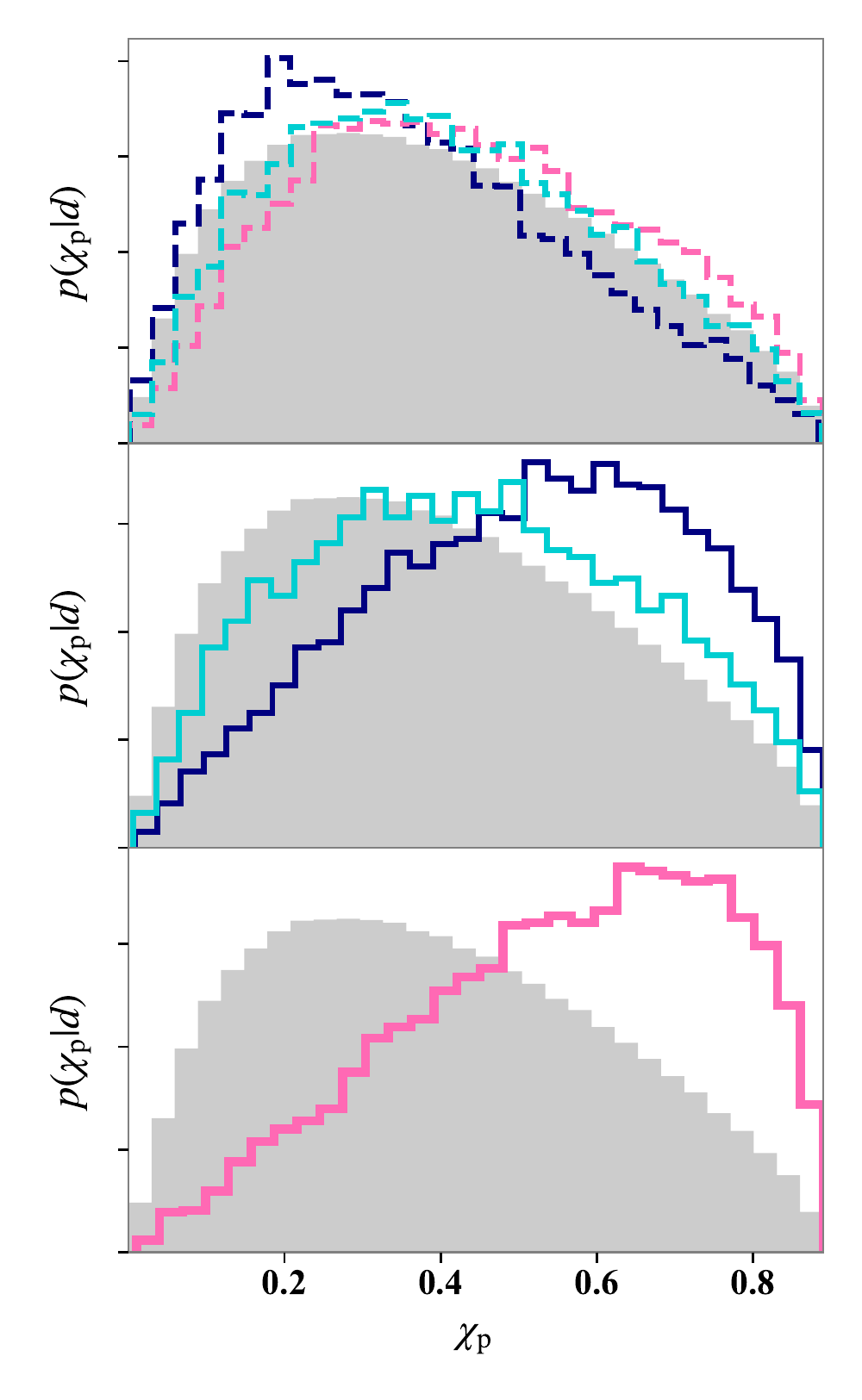}
    \end{subfigure}
    \caption{Results of \seobnre and \imrphenomPv analysis of simulated data using \eventname-like injections. \textit{Left}: Posterior distributions for eccentricity at \tenhz for \eventname-like injection studies with varying $e_{10}$ and $\chi_p$, obtained using \seobnre. \textit{Right}: Posterior distribution for precession parameter $\chi_p$ for \eventname-like injection studies with varying $e_{10}$ and $\chi_p$, recovered using \imrphenomPv. The prior distributions are shown in gray. \label{fig:eccentric_precessing_injections}}
\end{figure*}

\section{Injection studies}
Ideally, one would analyze gravitational-wave signals using models that include both precession and eccentricity. 
\red{This would allow simultaneous measurements of $\chi_p$ and $e_{10}$, as well as illuminating the full extent of the degeneracy between the two parameters and how that degeneracy changes with mass.}
Unfortunately, such models do not yet exist\red{; see \cite{Healy2009, Levin2011} for a theoretical background of eccentric and precessing binary dynamics and waveforms}. 
Thus, we use numerical tests to explore how our limited waveform models affect what we infer about eccentricity and precession.
We generate six \eventname-like waveform templates using different waveform models, each with different values of $e_{10}$ and $\chi_p$; see Table \ref{tab:injected_recovered_values}. 
Other parameters are identical to those in Table \ref{tab:injected_values}. 
Using \bilby, we inject these waveforms into simulated detector networks consisting of LIGO Hanford, LIGO Livingston, and Virgo, with noise power spectral densities matching those used for analysis of \eventname~\citep{GW190521}. 
For each injection, we recover the signal using both the aligned-spin eccentric model \seobnre and the quasi-circular precessing model \imrphenomPv.

In Fig.~\ref{fig:eccentric_precessing_injections}, we compare the posterior distributions for $e_{10}$ (obtained using \seobnre) and $\chi_p$ (obtained using \imrphenomPv) for all injections. 
When circular, non-precessing waveforms are injected, the \seobnre analysis recovers posterior distributions for $e_{10}$ consistent with the prior below the 90\% credible upper limit, $e_{10}^\mathrm{max} \leq 0.025$ ($0.032$, $0.055$) for injected \imrphenomD (\nrsurfour, \seobnre) waveforms. For these same waveforms, \imrphenomPv analysis recovers posteriors consistent with the prior on $\chi_p$.
When we increase only $\chi_p$, the posteriors on both $\chi_p$ and $e_{10}$ skew away from their priors. This is most notable for the \nrsurfour injections, suggesting that higher-order modes (included in \nrsurfour, but not in \imrphenomPv) may be important for distinguishing precession and eccentricity.
When we increase $e_{10}$, both posteriors deviate from their priors, more significantly than for the increased-precession case.
These injection studies demonstrate that, for \eventname-like binaries, precession may be mistaken for eccentricity, and that the imprint of eccentricity may be mistaken for that of precession.
We provide the full posterior distributions for all parameters in these injection studies online.\footnote{\href{https://git.ligo.org/isobel.romero-shaw/gw190521.1/injection_studies}{git.ligo.org/isobel.romero-shaw/gw190521.1/injection\_studies}}

\section{Discussion}
Assuming the aligned-spin \seobnre waveform model, we infer an eccentricity $e_{10} \gtrsim 0.1$ for \eventname. 
We find that the \seobnre waveform is slightly preferred over the circular-waveform models \nrsurfour and \imrphenomPv, both of which allow for precession.
While we lack a waveform model that can simultaneously account for precession and eccentricity, \eventname could be later verified as the first detection of a binary black hole with $e_{10} \geq 0.1$.
The presence of either precession or eccentricity adds weight to the hypothesis that the progenitor of \eventname formed dynamically.

\cite{Samsing17} predicts there are $\sim 19$ dynamical mergers with $e_{10} < 0.1$ for every merger with $e_{10} \geq 0.1$---a prediction thought to be robust to details about the globular cluster model; see also~\cite{Rodriguez18a} \red{and \cite{Martinez2020}}.
From the public alerts listed on GraceDB\footnote{\href{https://gracedb.ligo.org/superevents/public/O3/}{gracedb.ligo.org/superevents/public/O3/}}, there are $\mathcal{O}(30)$ binary black hole mergers from the first half of LIGO--Virgo's third observing run (O3a).
Combining these with the results of \cite{GWTC-1} and \cite{IAS-0, IAS-1, IAS-2}, the total number of binary black holes observed in gravitational waves is $\mathcal{O}(50)$.
If globular cluster mergers dominate LIGO and Virgo's observed black hole mergers, we expect $2.5_{-2.5}^{+2.0}$ mergers with $e_{10} \geq 0.1$ from the first $50$ binary black hole observations.
Thus, it would not be surprising if \eventname is determined to be highly eccentric.
Moreover, if \eventname is eccentric, then O3a may provide us with another $1.5_{-1.5}^{+2.0}$ events with $e_{10} \geq 0.1$\red{, assuming that O3a searches did not miss them; the signals of highly eccentric binaries may be missed by CBC and burst searches~\citep{East2013}}.

We note that while \eventname may have formed within a globular cluster, this is not its only viable formation pathway. Dynamical formation may also occur in active galactic nuclei~\citep[e.g.,][]{KocsisLevin2012, Yang19, Grobner20}, nuclear star clusters~\citep[e.g.,][]{Mapelli20}, young open clusters~\citep[e.g.,][]{DiCarlo19} \red{and young massive clusters~\citep{Banerjee2017, Banerjee2018a, Banerjee2018b, Banerjee2020, Kremer2020}}.
\red{
Mergers in young star clusters are likely to take place after ejection, giving the binary ample time to circularise and making young star clusters a less promising explanation for eccentric binaries. 
Both active galactic nuclei and globular clusters may produce binary black holes with misaligned spin and/or eccentricity, and so it is not clear which dynamical formation pathway is favoured for \eventname.
Regardless, both precession and eccentricity are signatures of dynamical formation; therefore, \eventname is likely to have formed in a dense stellar environment conducive to dynamical interactions}. 

\red{In dense environments like those mentioned above, binary black hole merger remnants may have masses within the mass gap. 
If these mergers are retained within the cluster, then they may merge again, producing intermediate-mass black holes. 
As an alternative to hierarchical black hole mergers, \cite{RoupasKazanas2020} argue that black holes may accrete enough gas in proto-clusters to enter into the mass gap. 
Another option is the direct collapse of stellar merger remnants to mass-gap black holes~\citep[e.g.,][]{Spera19, Kremer2020}.}
\red{These black holes} may undergo subsequent dynamical mergers if their environments are sufficiently densely populated. 
Although the high masses of \eventname render it incompatible with current models of isolated binary evolution, these masses can be produced in models where various model assumptions are substantially relaxed~\citep[see, e.g.,][]{Stevenson19, Farmer19, Marchant20}.
 
For \eventname-like signals, we highlight the degeneracy between eccentricity and precession\footnote{The degeneracy between eccentricity and precession is less pronounced for less massive systems, which have longer signals in-band.}.
This complements the results of \cite{JuanHeadOn}, who found that for the gravitational-wave signal of a head-on black-hole collision ($e_{10}=1$) with total mass in the range $M \in \unit[(130, 300)]{M_\odot}$ can be indistinguishable from the signal of a much more distant quasi-circular precessing binary.
Recently, a candidate electromagnetic counterpart for \eventname was observed at $\approx \unit[2.8]{Gpc}$ and reported \red{by \cite{EMCounterpart}, who propose that a binary black hole merger in an AGN disk might have such a counterpart}.
Extrapolating between the $e_{10}=1$ results from~\cite{JuanHeadOn} and the results shown here, the detected distance of \eventname in gravitational waves is consistent with the electromagnetic counterpart if \eventname had an eccentricity in the range $0.2 < e_{10} < 1.0$, a region of parameter space that cannot be fully explored with existing gravitational waveform models. However, new developments in eccentric waveforms~\citep[see, e.g.,][]{TEOBResumS} may allow us to start probing previously unexplored parameter space in the near future. 
\red{If the transient reported by \cite{EMCounterpart} is truly an electromagnetic counterpart emanating from an AGN disk merger, it would be consistent with the hypothesis that \eventname was an eccentric binary, since orbital eccentricity vastly increases the merger rate from such environments; see \cite{Grobner20}.} 

\textit{Note added.---}During the final stages of preparation of this manuscript, we became aware of the work of \cite{Gayathri}, who compare numerical-relativity waveform simulations to \eventname. Numerical relativity waveforms are too computationally expensive to be used for Bayesian parameter estimation. However, the fact that \cite{Gayathri} find that eccentric numerical-relativity simulations are consistent with \eventname supports the conclusions drawn in our work.

\section*{Acknowledgements}
\red{We thank the anonymous referee for their thoughtful suggestions, which improved the manuscript.} 
We thank Colm Talbot, Max Isi, Alan Weinstein, Tito Dal Canton, Christopher Berry and Chase Kimball for fruitful suggestions and illuminating discussions. We thank Rory Smith for assistance with parallel \bilby. This work is supported through Australian Research Council (ARC) Future Fellowships FT150100281, FT160100112, Centre of Excellence CE170100004, and Discovery Project DP180103155. JCB acknowledges support from the Direct Grant of the CUHK Research Committee, Project ID: 4053406
Computing was performed on the OzSTAR Australian national facility at Swinburne University of Technology, which receives funding in part from the Astronomy National Collaborative Research Infrastructure Strategy (NCRIS) allocation provided by the Australian Government, and the LIGO Laboratory computing cluster at California Institute of Technology, supported by National Science Foundation Grants PHY-0757058 and PHY-0823459.
LIGO was constructed by the California Institute of Technology and Massachusetts Institute of Technology with funding from the National Science Foundation and operates under cooperative agreement PHY-1764464. Virgo is funded by the French Centre National de Recherche Scientifique (CNRS), the Italian Istituto Nazionale della Fisica Nucleare (INFN) and the Dutch Nikhef, with contributions by Polish and Hungarian institutes.

\bibliography{main}

\begin{thebibliography}{}
\expandafter\ifx\csname natexlab\endcsname\relax\def\natexlab#1{#1}\fi
\providecommand{\url}[1]{\href{#1}{#1}}
\providecommand{\dodoi}[1]{doi:~\href{http://doi.org/#1}{\nolinkurl{#1}}}
\providecommand{\doeprint}[1]{\href{http://ascl.net/#1}{\nolinkurl{http://ascl.net/#1}}}
\providecommand{\doarXiv}[1]{\href{https://arxiv.org/abs/#1}{\nolinkurl{https://arxiv.org/abs/#1}}}

\bibitem[{Abbott {et~al.}(2020{\natexlab{a}})}]{GWOSC}
Abbott, {et~al.} 2020{\natexlab{a}}, Gravitational Wave Open Science Center
  Strain Data Release for GW190521, LIGO Open Science Center.
\newblock
  \url{www.gw-openscience.org/eventapi/html/O3_Discovery_Papers/GW190521/v2/}

\bibitem[{{Abbott} {et~al.}(2016{\natexlab{a}})}]{abbott16_gw150914_detection}
{Abbott}, B.~P., {et~al.} 2016{\natexlab{a}}, \prl, 116, 061102,
  \dodoi{10.1103/PhysRevLett.116.061102}

\bibitem[{{Abbott} {et~al.}(2016{\natexlab{b}})}]{abbott16_01BBH}
---. 2016{\natexlab{b}}, Phys. Rev. X, 6, 041015,
  \dodoi{10.1103/PhysRevX.6.041015}

\bibitem[{Abbott {et~al.}(2018)}]{Aasi13}
Abbott, B.~P., {et~al.} 2018, Living Rev. Rel., 21, 3,
  \dodoi{10.1007/s41114-018-0012-9, 10.1007/lrr-2016-1}

\bibitem[{Abbott {et~al.}(2019{\natexlab{a}})}]{GWTC-1}
---. 2019{\natexlab{a}}, Phys. Rev. X, 9, 031040,
  \dodoi{10.1103/PhysRevX.9.031040}

\bibitem[{Abbott {et~al.}(2019{\natexlab{b}})}]{EccentricCWB19}
---. 2019{\natexlab{b}}, arXiv e-prints, arXiv:1907.09384.
\newblock \doarXiv{1907.09384}

\bibitem[{Abbott {et~al.}(2020{\natexlab{b}})Abbott, Abbott, Abraham, Acernese,
  Ackley, {et~al.}}]{GW190521}
Abbott, R., Abbott, T.~D., Abraham, S., {et~al.} 2020{\natexlab{b}}, Phys. Rev.
  Lett., 125, 101102, \dodoi{10.1103/PhysRevLett.125.101102}

\bibitem[{Abbott {et~al.}(2020{\natexlab{c}})Abbott, Abbott, Abraham, Acernese,
  Ackley, {et~al.}}]{GW190521_implications}
---. 2020{\natexlab{c}}, \dodoi{10.3847/2041-8213/aba493}

\bibitem[{Acernese {et~al.}(2015)}]{AdvancedVirgo}
Acernese, F., {et~al.} 2015, Class. Quant. Grav., 32, 024001,
  \dodoi{10.1088/0264-9381/32/2/024001}

\bibitem[{Antonini {et~al.}(2017)}]{Antonini17}
Antonini, F., {et~al.} 2017, Astrophys. J., 841, 77,
  \dodoi{10.3847/1538-4357/aa6f5e}

\bibitem[{Ashton {et~al.}(2019)}]{bilby}
Ashton, G., {et~al.} 2019, Astrophys. J. Suppl., 241, 27,
  \dodoi{10.3847/1538-4365/ab06fc}

\bibitem[{{Banerjee}(2017)}]{Banerjee2017}
{Banerjee}, S. 2017, \mnras, 467, 524, \dodoi{10.1093/mnras/stw3392}

\bibitem[{{Banerjee}(2018{\natexlab{a}})}]{Banerjee2018a}
---. 2018{\natexlab{a}}, \mnras, 473, 909, \dodoi{10.1093/mnras/stx2347}

\bibitem[{{Banerjee}(2018{\natexlab{b}})}]{Banerjee2018b}
---. 2018{\natexlab{b}}, \mnras, 481, 5123, \dodoi{10.1093/mnras/sty2608}

\bibitem[{{Banerjee}(2020)}]{Banerjee2020}
---. 2020, \mnras, \dodoi{10.1093/mnras/staa2392}

\bibitem[{{Belczynski}(2020)}]{Belczynski2020}
{Belczynski}, K. 2020, arXiv e-prints, arXiv:2009.13526.
\newblock \doarXiv{2009.13526}

\bibitem[{{Belczynski} {et~al.}(2016){Belczynski}, {Heger}, {Gladysz},
  {Ruiter}, {Woosley}, {Wiktorowicz}, {Chen}, {Bulik}, {O'Shaughnessy}, {Holz},
  {Fryer}, \& {Berti}}]{Belczynski16}
{Belczynski}, K., {Heger}, A., {Gladysz}, W., {et~al.} 2016, \aap, 594, A97,
  \dodoi{10.1051/0004-6361/201628980}

\bibitem[{Bethe \& Brown(1998)}]{Bethe98}
Bethe, H.~A., \& Brown, G.~E. 1998, Astrophys. J., 506, 780,
  \dodoi{10.1086/306265}

\bibitem[{Bouffanais {et~al.}(2019)}]{Bouffanais19}
Bouffanais, Y., {et~al.} 2019.
\newblock \doarXiv{1905.11054}

\bibitem[{Bustillo {et~al.}(2020)Bustillo, Sanchis-Gual, Torres-Forné, \&
  Font}]{JuanHeadOn}
Bustillo, J.~C., Sanchis-Gual, N., Torres-Forné, A., \& Font, J.~A. 2020,
  Confusing head-on and precessing intermediate-mass binary black hole mergers

\bibitem[{Cao \& Han(2017)}]{SEOBNRE}
Cao, Z., \& Han, W.-B. 2017, Phys. Rev., D96, 044028,
  \dodoi{10.1103/PhysRevD.96.044028}

\bibitem[{Chatziioannou {et~al.}(2019)}]{Chatziioannou2019}
Chatziioannou, K., {et~al.} 2019, Phys. Rev. D, 100, 104015

\bibitem[{{Chiaramello} \& {Nagar}(2020)}]{TEOBResumS}
{Chiaramello}, D., \& {Nagar}, A. 2020, \prd, 101, 101501,
  \dodoi{10.1103/PhysRevD.101.101501}

\bibitem[{de~Mink \& Mandel(2016)}]{deMink16}
de~Mink, S.~E., \& Mandel, I. 2016, Mon. Not. Roy. Astron. Soc., 460, 3545,
  \dodoi{10.1093/mnras/stw1219}

\bibitem[{{de Mink} {et~al.}(2010)}]{deMink10}
{de Mink}, S.~E., {et~al.} 2010, in American Institute of Physics Conference
  Series, Vol. 1314, American Institute of Physics Conference Series, ed.
  V.~{Kalogera} \& M.~{van der Sluys}, 291--296

\bibitem[{{Di Carlo} {et~al.}(2019{\natexlab{a}}){Di Carlo}, {Giacobbo},
  {Mapelli}, {Pasquato}, {Spera}, {Wang}, \& {Haardt}}]{DiCarlo2019}
{Di Carlo}, U.~N., {Giacobbo}, N., {Mapelli}, M., {et~al.} 2019{\natexlab{a}},
  \mnras, 487, 2947, \dodoi{10.1093/mnras/stz1453}

\bibitem[{{Di Carlo} {et~al.}(2019{\natexlab{b}}){Di Carlo}, {Giacobbo},
  {Mapelli}, {Pasquato}, {Spera}, {Wang}, \& {Haardt}}]{DiCarlo19}
---. 2019{\natexlab{b}}, \mnras, 487, 2947, \dodoi{10.1093/mnras/stz1453}

\bibitem[{{East} {et~al.}(2013){East}, {McWilliams}, {Levin}, \&
  {Pretorius}}]{East2013}
{East}, W.~E., {McWilliams}, S.~T., {Levin}, J., \& {Pretorius}, F. 2013, \prd,
  87, 043004, \dodoi{10.1103/PhysRevD.87.043004}

\bibitem[{{Farmer} {et~al.}(2019){Farmer}, {Renzo}, {de Mink}, {Marchant}, \&
  {Justham}}]{Farmer19}
{Farmer}, R., {Renzo}, M., {de Mink}, S.~E., {Marchant}, P., \& {Justham}, S.
  2019, \apj, 887, 53, \dodoi{10.3847/1538-4357/ab518b}

\bibitem[{Farr {et~al.}(2017)}]{Farr17}
Farr, W.~M., {et~al.} 2017, Nature, 548, 426, \dodoi{10.1038/nature23453}

\bibitem[{Fishbach \& Holz(2017)}]{Fishbach17}
Fishbach, M., \& Holz, D.~E. 2017, Astrophys. J., 851, L25,
  \dodoi{10.3847/2041-8213/aa9bf6}

\bibitem[{{Fishbach} \& {Holz}(2020)}]{Fishbach2020}
{Fishbach}, M., \& {Holz}, D.~E. 2020, arXiv e-prints, arXiv:2009.05472.
\newblock \doarXiv{2009.05472}

\bibitem[{Fishbach {et~al.}(2017)}]{Fishbach17a}
Fishbach, M., {et~al.} 2017, Astrophys. J., 840, L24,
  \dodoi{10.3847/2041-8213/aa7045}

\bibitem[{Fragione \& Bromberg(2019)}]{Fragione19b}
Fragione, G., \& Bromberg, O. 2019, arXiv e-prints, arXiv:1903.09659.
\newblock \doarXiv{1903.09659}

\bibitem[{Fragione \& Kocsis(2018)}]{Fragione18}
Fragione, G., \& Kocsis, B. 2018, Phys. Rev. Lett., 121,
  \dodoi{10.1103/PhysRevLett.121.161103}

\bibitem[{{Fragione} \& {Kocsis}(2019{\natexlab{a}})}]{triplespin}
{Fragione}, G., \& {Kocsis}, B. 2019{\natexlab{a}}, arXiv e-prints,
  arXiv:1910.00407.
\newblock \doarXiv{1910.00407}

\bibitem[{{Fragione} \& {Kocsis}(2019{\natexlab{b}})}]{quadruples2}
---. 2019{\natexlab{b}}, \mnras, 486, 4781, \dodoi{10.1093/mnras/stz1175}

\bibitem[{{Fragione} {et~al.}(2020){Fragione}, {Loeb}, \&
  {Rasio}}]{Fragione2020}
{Fragione}, G., {Loeb}, A., \& {Rasio}, F.~A. 2020, arXiv e-prints,
  arXiv:2009.05065.
\newblock \doarXiv{2009.05065}

\bibitem[{Fragione {et~al.}(2020)}]{fragione20}
Fragione, G., {et~al.} 2020.
\newblock \doarXiv{2002.11278}

\bibitem[{{Gayathri} {et~al.}(2020){Gayathri}, {Healy}, {Lange}, {O'Brien},
  {Szczepanczyk}, {Bartos}, {Campanelli}, {Klimenko}, {Lousto}, \&
  {O'Shaughnessy}}]{Gayathri}
{Gayathri}, V., {Healy}, J., {Lange}, J., {et~al.} 2020, arXiv e-prints,
  arXiv:2009.05461.
\newblock \doarXiv{2009.05461}

\bibitem[{Gerosa \& Berti(2017)}]{Gerosa17}
Gerosa, D., \& Berti, E. 2017, Phys. Rev., D95, 124046,
  \dodoi{10.1103/PhysRevD.95.124046}

\bibitem[{Gond\'an \& Kocsis(2019)}]{Gondan18}
Gond\'an, L., \& Kocsis, B. 2019, Astrophys. J., 871, 178,
  \dodoi{10.3847/1538-4357/aaf893}

\bibitem[{Gond\'an {et~al.}(2018)}]{Gondan17}
Gond\'an, L., {et~al.} 2018, Astrophys. J., 860, 5,
  \dodoi{10.3847/1538-4357/aabfee}

\bibitem[{{Graham} {et~al.}(2020){Graham}, {Ford}, {McKernan}, {Ross}, {Stern},
  {Burdge}, {Coughlin}, {Djorgovski}, {Drake}, {Duev}, {Kasliwal}, {Mahabal},
  {van Velzen}, {Belecki}, {Bellm}, {Burruss}, {Cenko}, {Cunningham}, {Helou},
  {Kulkarni}, {Masci}, {Prince}, {Reiley}, {Rodriguez}, {Rusholme}, {Smith}, \&
  {Soumagnac}}]{EMCounterpart}
{Graham}, M.~J., {Ford}, K.~E.~S., {McKernan}, B., {et~al.} 2020, \prl, 124,
  251102, \dodoi{10.1103/PhysRevLett.124.251102}

\bibitem[{{Gr{\"o}bner} {et~al.}(2020){Gr{\"o}bner}, {Ishibashi}, {Tiwari},
  {Haney}, \& {Jetzer}}]{Grobner20}
{Gr{\"o}bner}, M., {Ishibashi}, W., {Tiwari}, S., {Haney}, M., \& {Jetzer}, P.
  2020, \aap, 638, A119, \dodoi{10.1051/0004-6361/202037681}

\bibitem[{{Healy} {et~al.}(2009){Healy}, {Levin}, \& {Shoemaker}}]{Healy2009}
{Healy}, J., {Levin}, J., \& {Shoemaker}, D. 2009, \prl, 103, 131101,
  \dodoi{10.1103/PhysRevLett.103.131101}

\bibitem[{{Heger} \& {Woosley}(2002)}]{HegerWoosley02}
{Heger}, A., \& {Woosley}, S.~E. 2002, Astrophy. J., 567, 532,
  \dodoi{10.1086/338487}

\bibitem[{Hinder {et~al.}(2008)}]{Hinder07}
Hinder, I., {et~al.} 2008, Phys. Rev. D, D77, 081502,
  \dodoi{10.1103/PhysRevD.77.081502}

\bibitem[{{Ivanova} {et~al.}(2013)}]{Ivanova13}
{Ivanova}, N., {et~al.} 2013, \aapr, 21, 59, \dodoi{10.1007/s00159-013-0059-2}

\bibitem[{Khan {et~al.}(2016)}]{Khan15}
Khan, S., {et~al.} 2016, Phys. Rev., D93, 044007,
  \dodoi{10.1103/PhysRevD.93.044007}

\bibitem[{Kimball {et~al.}(2019)Kimball, Berry, \& Kalogera}]{Kimball2019}
Kimball, C., Berry, C., \& Kalogera, V. 2019, RNAAS, 4, 2

\bibitem[{Kimball {et~al.}(2020)}]{Kimball2020}
Kimball, C., {et~al.} 2020

\bibitem[{{Kocsis} \& {Levin}(2012)}]{KocsisLevin2012}
{Kocsis}, B., \& {Levin}, J. 2012, \prd, 85, 123005,
  \dodoi{10.1103/PhysRevD.85.123005}

\bibitem[{{Kozai}(1962)}]{Kozai62}
{Kozai}, Y. 1962, Astrophys. J., 67, 591, \dodoi{10.1086/108790}

\bibitem[{{Kremer} {et~al.}(2020){Kremer}, {Spera}, {Becker}, {Chatterjee}, {Di
  Carlo}, {Fragione}, {Rodriguez}, {Ye}, \& {Rasio}}]{Kremer2020}
{Kremer}, K., {Spera}, M., {Becker}, D., {et~al.} 2020, arXiv e-prints,
  arXiv:2006.10771.
\newblock \doarXiv{2006.10771}

\bibitem[{Kruckow {et~al.}(2016)}]{Kruckow16}
Kruckow, M.~U., {et~al.} 2016, Astron. Astrophys., 596, A58,
  \dodoi{10.1051/0004-6361/201629420}

\bibitem[{{Levin} {et~al.}(2011){Levin}, {McWilliams}, \&
  {Contreras}}]{Levin2011}
{Levin}, J., {McWilliams}, S.~T., \& {Contreras}, H. 2011, Classical and
  Quantum Gravity, 28, 175001, \dodoi{10.1088/0264-9381/28/17/175001}

\bibitem[{{Lidov}(1962)}]{Lidov62}
{Lidov}, M.~L. 1962, Planetary and Space Science, 9, 719,
  \dodoi{10.1016/0032-0633(62)90129-0}

\bibitem[{{Liu} \& {Lai}(2019)}]{quadruples}
{Liu}, B., \& {Lai}, D. 2019, \mnras, 483, 4060, \dodoi{10.1093/mnras/sty3432}

\bibitem[{Liu {et~al.}(2019)}]{Liu19}
Liu, B., {et~al.} 2019.
\newblock \doarXiv{1905.00427}

\bibitem[{{Liu} {et~al.}(2019)}]{validationSEOBNRE}
{Liu}, X., {et~al.} 2019, arXiv e-prints, arXiv:1910.00784.
\newblock \doarXiv{1910.00784}

\bibitem[{{Livio} \& {Soker}(1988)}]{Livio88}
{Livio}, M., \& {Soker}, N. 1988, Astrophys. J., 329, 764,
  \dodoi{10.1086/166419}

\bibitem[{Lower {et~al.}(2018)}]{Lower18}
Lower, M., {et~al.} 2018, Phys. Rev. D, 98, \dodoi{10.1103/PhysRevD.98.083028}

\bibitem[{{Mapelli} {et~al.}(2020){Mapelli}, {Santoliquido}, {Bouffanais},
  {Arca Sedda}, {Giacobbo}, {Artale}, \& {Ballone}}]{Mapelli20}
{Mapelli}, M., {Santoliquido}, F., {Bouffanais}, Y., {et~al.} 2020, arXiv
  e-prints, arXiv:2007.15022.
\newblock \doarXiv{2007.15022}

\bibitem[{{Marchant} {et~al.}(2016){Marchant}, {Langer}, {Podsiadlowski},
  {Tauris}, \& {Moriya}}]{Marchant16}
{Marchant}, P., {Langer}, N., {Podsiadlowski}, P., {Tauris}, T.~M., \&
  {Moriya}, T.~J. 2016, \aap, 588, A50, \dodoi{10.1051/0004-6361/201628133}

\bibitem[{{Marchant} \& {Moriya}(2020)}]{Marchant20}
{Marchant}, P., \& {Moriya}, T. 2020, arXiv e-prints, arXiv:2007.06220.
\newblock \doarXiv{2007.06220}

\bibitem[{{Martinez} {et~al.}(2020){Martinez}, {Fragione}, {Kremer},
  {Chatterjee}, {Rodriguez}, {Samsing}, {Ye}, {Weatherford}, {Zevin}, {Naoz},
  \& {Rasio}}]{Martinez2020}
{Martinez}, M. A.~S., {Fragione}, G., {Kremer}, K., {et~al.} 2020, arXiv
  e-prints, arXiv:2009.08468.
\newblock \doarXiv{2009.08468}

\bibitem[{{McKernan} {et~al.}(2020){McKernan}, {Ford}, \&
  {O'Shaughnessy}}]{McKernan2020}
{McKernan}, B., {Ford}, K.~E.~S., \& {O'Shaughnessy}, R. 2020, arXiv e-prints,
  arXiv:2002.00046.
\newblock \doarXiv{2002.00046}

\bibitem[{{Morscher} {et~al.}(2015)}]{Morscher15}
{Morscher}, M., {et~al.} 2015, \apj, 800, 9, \dodoi{10.1088/0004-637X/800/1/9}

\bibitem[{O'Leary {et~al.}(2006)}]{OLeary05}
O'Leary, R.~M., {et~al.} 2006, Astrophys. J., 637, 937, \dodoi{10.1086/498446}

\bibitem[{{{\"O}zel} {et~al.}(2010){{\"O}zel}, {Psaltis}, {Narayan}, \&
  {McClintock}}]{Ozel10}
{{\"O}zel}, F., {Psaltis}, D., {Narayan}, R., \& {McClintock}, J.~E. 2010,
  \apj, 725, 1918, \dodoi{10.1088/0004-637X/725/2/1918}

\bibitem[{{Payne} {et~al.}(2019){Payne}, {Talbot}, \& {Thrane}}]{Payne}
{Payne}, E., {Talbot}, C., \& {Thrane}, E. 2019, \prd, 100, 123017,
  \dodoi{10.1103/PhysRevD.100.123017}

\bibitem[{Peters(1964)}]{Peters64}
Peters, P.~C. 1964, Phys. Rev., 136, B1224, \dodoi{10.1103/PhysRev.136.B1224}

\bibitem[{Portegies~Zwart \& McMillan(2000)}]{PortegiesZwart99}
Portegies~Zwart, S.~F., \& McMillan, S. 2000, Astrophys. J., 528, L17,
  \dodoi{10.1086/312422}

\bibitem[{Randall \& Xianyu(2018{\natexlab{a}})}]{Randall17}
Randall, L., \& Xianyu, Z.-Z. 2018{\natexlab{a}}, Astrophys. J., 853,
  \dodoi{10.3847/1538-4357/aaa1a2}

\bibitem[{Randall \& Xianyu(2018{\natexlab{b}})}]{Randall18}
---. 2018{\natexlab{b}}, Astrophys. J., 864, 134,
  \dodoi{10.3847/1538-4357/aad7fe}

\bibitem[{{Rastello} {et~al.}(2020){Rastello}, {Mapelli}, {Di Carlo},
  {Giacobbo}, {Santoliquido}, {Spera}, {Ballone}, \& {Iorio}}]{Rastello2020}
{Rastello}, S., {Mapelli}, M., {Di Carlo}, U.~N., {et~al.} 2020, \mnras,
  \dodoi{10.1093/mnras/staa2018}

\bibitem[{Rodriguez \& Antonini(2018)}]{Rodriguez18jqu}
Rodriguez, C.~L., \& Antonini, F. 2018, Astrophys. J., 863, 7,
  \dodoi{10.3847/1538-4357/aacea4}

\bibitem[{Rodriguez {et~al.}(2016)}]{Rodriguez16}
Rodriguez, C.~L., {et~al.} 2016, Astrophys. J., 832, L2,
  \dodoi{10.3847/2041-8205/832/1/L2}

\bibitem[{Rodriguez {et~al.}(2018{\natexlab{a}})}]{Rodriguez18b}
---. 2018{\natexlab{a}}, Phys. Rev., D98, 123005,
  \dodoi{10.1103/PhysRevD.98.123005}

\bibitem[{Rodriguez {et~al.}(2018{\natexlab{b}})}]{Rodriguez18a}
---. 2018{\natexlab{b}}, Phys. Rev. Lett., 120, 151101,
  \dodoi{10.1103/PhysRevLett.120.151101}

\bibitem[{Rodriguez {et~al.}(2019)}]{Rodriguez19}
---. 2019.
\newblock \doarXiv{1906.10260}

\bibitem[{{Romero-Shaw} {et~al.}(2019){Romero-Shaw}, {Lasky}, \&
  {Thrane}}]{RoSho19}
{Romero-Shaw}, I.~M., {Lasky}, P.~D., \& {Thrane}, E. 2019, \mnras, 490, 5210,
  \dodoi{10.1093/mnras/stz2996}

\bibitem[{{Romero-Shaw} {et~al.}(2020){Romero-Shaw}, {Talbot}, {Biscoveanu},
  {D'Emilio}, {Ashton}, {Berry}, {Coughlin}, {Galaudage}, {Hoy}, {Huebner},
  {Phukon}, {Pitkin}, {Rizzo}, {Sarin}, {Smith}, {Stevenson}, {Vajpeyi},
  {Arene}, {Athar}, {Banagiri}, {Bose}, {Carney}, {Chatziioannou}, {Cotesta},
  {Edelman}, {Garcia-Quiros}, {Ghosh}, {Green}, {Haster}, {Kim},
  {Hernandez-Vivanco}, {Magana Hernandez}, {Karathanasis}, {Lasky}, {De Lillo},
  {Lower}, {Macleod}, {Mateu-Lucena}, {Miller}, {Millhouse}, {Morisaki}, {Oh},
  {Ossokine}, {Payne}, {Powell}, {Puerrer}, {Ramos-Buades}, {Raymond},
  {Thrane}, {Veitch}, {Williams}, {Williams}, \& {Xiao}}]{bilbyGWTC1}
{Romero-Shaw}, I.~M., {Talbot}, C., {Biscoveanu}, S., {et~al.} 2020, arXiv
  e-prints, arXiv:2006.00714.
\newblock \doarXiv{2006.00714}

\bibitem[{{Roupas} \& {Kazanas}(2019)}]{RoupasKazanas2020}
{Roupas}, Z., \& {Kazanas}, D. 2019, \aap, 632, L8,
  \dodoi{10.1051/0004-6361/201937002}

\bibitem[{{Sakstein} {et~al.}(2020){Sakstein}, {Croon}, {McDermott},
  {Straight}, \& {Baxter}}]{Sakstein2020}
{Sakstein}, J., {Croon}, D., {McDermott}, S.~D., {Straight}, M.~C., \&
  {Baxter}, E.~J. 2020, arXiv e-prints, arXiv:2009.01213.
\newblock \doarXiv{2009.01213}

\bibitem[{Samsing(2018)}]{Samsing17}
Samsing, J. 2018, Phys. Rev. D, D97, 103014, \dodoi{10.1103/PhysRevD.97.103014}

\bibitem[{Samsing \& D'Orazio(2018)}]{SamsingDOrazio18}
Samsing, J., \& D'Orazio, D.~J. 2018, Mon. Not. Roy. Astron. Soc., 481,
  \dodoi{10.1093/mnras/sty2334}

\bibitem[{Samsing \& Ramirez-Ruiz(2017)}]{SamsingRamirez17}
Samsing, J., \& Ramirez-Ruiz, E. 2017, Astrophys. J., 840, L14,
  \dodoi{10.3847/2041-8213/aa6f0b}

\bibitem[{Samsing {et~al.}(2014)}]{Samsing13}
Samsing, J., {et~al.} 2014, Astrophys. J., 784, 71,
  \dodoi{10.1088/0004-637X/784/1/71}

\bibitem[{Samsing {et~al.}(2018)}]{Samsing18}
---. 2018, arXiv e-prints, arXiv:1802.08654.
\newblock \doarXiv{1802.08654}

\bibitem[{{Schmidt} {et~al.}(2012)}]{IMRPhenomP}
{Schmidt}, P., {et~al.} 2012, \prd, 86, 104063

\bibitem[{Sigurdsson \& Hernquist(1993)}]{Sigurdsson93}
Sigurdsson, S., \& Hernquist, L. 1993, Nature, 364, 423,
  \dodoi{10.1038/364423a0}

\bibitem[{Silsbee \& Tremaine(2017)}]{Silsbee16}
Silsbee, K., \& Tremaine, S. 2017, Astrophys. J., 836, 39,
  \dodoi{10.3847/1538-4357/aa5729}

\bibitem[{{Smith} {et~al.}(2019){Smith}, {Ashton}, {Vajpeyi}, \&
  {Talbot}}]{ParallelBilby}
{Smith}, R., {Ashton}, G., {Vajpeyi}, A., \& {Talbot}, C. 2019, arXiv e-prints,
  arXiv:1909.11873.
\newblock \doarXiv{1909.11873}

\bibitem[{{Spera} {et~al.}(2019){Spera}, {Mapelli}, {Giacobbo}, {Trani},
  {Bressan}, \& {Costa}}]{Spera19}
{Spera}, M., {Mapelli}, M., {Giacobbo}, N., {et~al.} 2019, \mnras, 485, 889,
  \dodoi{10.1093/mnras/stz359}

\bibitem[{{Stevenson} {et~al.}(2019){Stevenson}, {Sampson}, {Powell},
  {Vigna-G{\'o}mez}, {Neijssel}, {Sz{\'e}csi}, \& {Mandel}}]{Stevenson19}
{Stevenson}, S., {Sampson}, M., {Powell}, J., {et~al.} 2019, \apj, 882, 121,
  \dodoi{10.3847/1538-4357/ab3981}

\bibitem[{Stevenson {et~al.}(2015)}]{Stevenson15bqa}
Stevenson, S., {et~al.} 2015, Astrophys. J., 810, 58,
  \dodoi{10.1088/0004-637X/810/1/58}

\bibitem[{Talbot \& Thrane(2017)}]{TalbotThrane17}
Talbot, C., \& Thrane, E. 2017, Phys. Rev. D, 96, 023012

\bibitem[{{Venumadhav} {et~al.}(2020){Venumadhav}, {Zackay}, {Roulet}, {Dai},
  \& {Zaldarriaga}}]{IAS-2}
{Venumadhav}, T., {Zackay}, B., {Roulet}, J., {Dai}, L., \& {Zaldarriaga}, M.
  2020, \prd, 101, 083030, \dodoi{10.1103/PhysRevD.101.083030}

\bibitem[{{Venumadhav} {et~al.}(2019)}]{IAS-0}
{Venumadhav}, T., {et~al.} 2019, \prd, 100, 023011,
  \dodoi{10.1103/PhysRevD.100.023011}

\bibitem[{Vitale {et~al.}(2017)}]{Vitale15}
Vitale, S., {et~al.} 2017, Class. Quant. Grav., 34, 03LT01,
  \dodoi{10.1088/1361-6382/aa552e}

\bibitem[{{Woosley}(2017)}]{Woosley17}
{Woosley}, S.~E. 2017, \apj, 836, 244, \dodoi{10.3847/1538-4357/836/2/244}

\bibitem[{{Yang} {et~al.}(2019){Yang}, {Bartos}, {Haiman}, {Kocsis},
  {M{\'a}rka}, {Stone}, \& {M{\'a}rka}}]{Yang2019}
{Yang}, Y., {Bartos}, I., {Haiman}, Z., {et~al.} 2019, \apj, 876, 122,
  \dodoi{10.3847/1538-4357/ab16e3}

\bibitem[{Yang {et~al.}(2019)Yang, Bartos, Gayathri, Ford, Haiman, Klimenko,
  Kocsis, M\'arka, M\'arka, McKernan, \& O'Shaughnessy}]{Yang19}
Yang, Y., Bartos, I., Gayathri, V., {et~al.} 2019, Phys. Rev. Lett., 123,
  181101, \dodoi{10.1103/PhysRevLett.123.181101}

\bibitem[{{Zackay} {et~al.}(2019){Zackay}, {Dai}, {Venumadhav}, {Roulet}, \&
  {Zaldarriaga}}]{IAS-1}
{Zackay}, B., {Dai}, L., {Venumadhav}, T., {Roulet}, J., \& {Zaldarriaga}, M.
  2019, arXiv e-prints, arXiv:1910.09528.
\newblock \doarXiv{1910.09528}

\bibitem[{Zevin {et~al.}(2017)}]{Zevin17}
Zevin, M., {et~al.} 2017, Astrophys. J., 846, 82,
  \dodoi{10.3847/1538-4357/aa8408}

\bibitem[{Zevin {et~al.}(2019{\natexlab{a}})}]{Zevin18}
---. 2019{\natexlab{a}}, Astrophys. J., 871, 91,
  \dodoi{10.3847/1538-4357/aaf6ec}

\bibitem[{Zevin {et~al.}(2019{\natexlab{b}})}]{Zevin19ns}
---. 2019{\natexlab{b}}.
\newblock \doarXiv{1906.11299}

\bibitem[{{Ziosi} {et~al.}(2014){Ziosi}, {Mapelli}, {Branchesi}, \&
  {Tormen}}]{Ziosi2014}
{Ziosi}, B.~M., {Mapelli}, M., {Branchesi}, M., \& {Tormen}, G. 2014, Monthly
  Notices of the Royal Astronomical Society, 441, 3703

\end{thebibliography}

\end{document}